\documentclass[report]{IEEEtran}
\IEEEoverridecommandlockouts
\usepackage{amsmath,amssymb,amsfonts}
\usepackage{algorithm}
\usepackage{algpseudocode}
\usepackage{array}

\usepackage{textcomp}
\usepackage{stfloats}
\usepackage{url}
\usepackage{verbatim}
\usepackage{cite}
\usepackage{gensymb}
\usepackage{xcolor}

\usepackage[T1]{fontenc}
\usepackage{helvet}

\usepackage[]{mdframed}

\newcommand{\framedtext}[1]{%
\par%
\noindent\fbox{%
    \parbox{\dimexpr\linewidth-2\fboxsep-2\fboxrule}{#1}%
}%
}

\usepackage{graphicx}    % Required for including images
\usepackage{caption}
\usepackage{subcaption}
\usepackage[acronym,nomain,nonumberlist]{glossaries}

\newcommand{\ba}{\mathbf{a}}
\newcommand{\bb}{\mathbf{b}}

\newcommand{\bj}{\mathbf{j}}

\newcommand{\bs}{\mathbf{s}}

\newcommand{\bw}{\mathbf{w}}
\newcommand{\bx}{\mathbf{x}}
\newcommand{\by}{\mathbf{y}}
\newcommand{\bz}{\mathbf{z}}

\newcommand{\bA}{\mathbf{A}}
\newcommand{\bB}{\mathbf{B}}

\newcommand{\bE}{\mathbf{E}}
\newcommand{\bF}{\mathbf{F}}

\newcommand{\bI}{\mathbf{I}}
\newcommand{\bJ}{\mathbf{J}}

\newcommand{\bQ}{\mathbf{Q}}
\newcommand{\bR}{\mathbf{R}}
\newcommand{\bS}{\mathbf{S}}

\newcommand{\bW}{\mathbf{W}}
\newcommand{\bX}{\mathbf{X}}
\newcommand{\bY}{\mathbf{Y}}

\newacronym{5G}{5G}{fifth-generation}
\newacronym{BS}{BS}{base station}
\newacronym{DMA}{DMA}{dynamic metasurface antenna}
\newacronym{RF}{RF}{radio-frequency}
\newacronym{AO}{AO}{alternative optimization}
\newacronym{ISAC}{ISAC}{Integrated sensing and communication}
\newacronym{MIMO}{MIMO}{multiple-input multiple-outpuT}
\newacronym{LIS}{LIS}{large intelligent surface}
\newacronym{MUI}{MUI}{multi-user interference}
\newacronym{RIS}{RIS}{reconfigurable intelligent surface}
\newacronym{MSE}{MSE}{mean square error}
\newacronym{RHS}{RHS}{reconfigurable holographic surfac}
\newacronym{CRLB}{CRLB}{Cramer-Rao lower bound}
\newacronym{SINR}{SINR}{signal-to-interference-plus-noise ratio}
\newacronym{gNB}{gNB}{gNodeB}
\newacronym{OFDM}{OFDM}{orthogonal frequency-division multiplexing}
\newacronym{6G}{6G}{sixth-generation}
\newacronym{AWGN}{AWGN}{additive white Gaussian noise}
\newacronym{QCQP}{QCQP}{quadratically constrained quadratic program}
\newacronym{SDR}{SDR}{semidefinite relaxation}
\newacronym{LoS}{LoS}{line-of-sight}
\newacronym{PSD}{PSD}{positive semidefinite}
\newacronym{FD}{FD}{full duplex}
\newacronym{SNR}{SNR}{signal-to-noise ratio}

\begin{document}

\title{Mathematical Approach in Hybrid Beamforming for ISAC Systems}

\author{Keivan Khosroshahi, and Michael Suppa
        % <-this % stops a space
\thanks{K. Khosroshahi is with Université Paris-Saclay, CNRS,
CentraleSupélec, Laboratoire des Signaux et Systémes, 91192 Gif-sur-Yvette,
France (keivan.khosroshahi@centralesupelec.fr). M. Suppa is with Roboception GmbH, 81241 Munich, Germany (michael.suppa@roboception.de). This work was supported by the European Commission through the H2020 MSCA 5GSmartFact
project under grant agreement 956670.}}

\maketitle

\begin{abstract}
This document serves as supplementary material for a journal paper submission, presenting detailed mathematical proofs and derivations that support the results outlined in the main manuscript. In this work, we formulate an \gls{ISAC} optimization problem aimed at fairly maximizing beampattern gains across multiple directions of interest, while ensuring that the \gls{SINR} requirements for communication users and the overall power constraint are satisfied. To solve this problem, we propose an optimization algorithm and provide a formal proof of its convergence.
\end{abstract}

\section{Introduction}
\gls{DMA} has recently been proposed as a practical implementation for transmitter and receiver intelligent metasurfaces \cite{smith2017analysis, shlezinger2021dynamic}. It consists of numerous artificial metasurface elements that can be dynamically reconfigured in real time, enabling flexible control over transmitted and received signals. This reconfigurability is achieved through the integration of switchable components within each element. \gls{DMA} elements are semi-passive and characterized by extremely low power consumption. Compared to large-scale antenna arrays employed in massive \gls{MIMO}, \gls{DMA} offers significantly lower power consumption and reduced hardware complexity, primarily due to its limited number of required \gls{RF} chains \cite{kimaryo2022downlink}. Additionally, the compact form factor of \gls{DMA} elements allows for the deployment of a greater number of elements relative to conventional massive \gls{MIMO} antennas. Signal processing techniques such as analog beamforming and antenna selection can also be performed directly within the \gls{DMA} architecture, eliminating the need for additional hardware typically required in hybrid systems \cite{kimaryo2022downlink}.

To date, \gls{DMA} has been primarily investigated in the domains of satellite communications, radar systems, and microwave imaging \cite{shlezinger2019dynamic}, where it has demonstrated notable advantages in terms of simplicity, energy efficiency, and flexibility. However, its integration into \gls{ISAC} systems—particularly in the context of hybrid beamforming—remains relatively underexplored. In this work, we formulate an \gls{ISAC} optimization problem aimed at fairly maximizing beampattern gains across multiple directions of interest, while ensuring that the \gls{SINR} requirements for communication users and overall power constraints are satisfied. To solve this problem, we propose an optimization algorithm and provide a formal proof of its convergence.

\section{Functions definition}
We define the parameters and functions that we use in this paper as follows:

\begin{align}
    \gamma_m &= \frac{\mathbf{h}_m \bb_m\bb_m^H \mathbf{h}_m^H}{\mathbf{h}_m \bB\bB^H
\mathbf{h}_m^H - \mathbf{h}_m \bb_m\bb_m^H \mathbf{h}_m^H + \sigma^2} \label{eq:SINR1} \\
    z(\theta, \phi) &= \ba(\theta, \phi)^T (\mathbf{Y_{s}} + \mathbf{Y_{ss}})^{-1}\mathbf{Y_{st}} (\mathbf{B}_c\mathbf{c}+\mathbf{B}_s\mathbf{s})\\
    P(\theta, \phi) &= \mathbb{E} (z(\theta, \phi)z^H(\theta, \phi)) 
    = \ba(\theta, \phi)^T (\mathbf{Y_{s}}  \notag\\
    &+ \mathbf{Y_{ss}})^{-1}\mathbf{Y_{st}} \bB\bB^H \mathbf{Y_{st}}^H (\mathbf{Y_{s}}^H + \mathbf{Y_{ss}}^H)^{-1} \ba(\theta, \phi)^* \label{eq:BP}\\
    P_{\text{tot}} &= \frac{1}{D}\sum_{d=1}^D P(\theta_d,\phi_d) \label{eq:BPtot}\\
    P_{t} &= \frac{1}{2} \text{Re}\left\{\text{tr}\{(\bY_{tt} - \bY_{st}^T(\bY_s + \bY_{ss})^{ - 1}\bY_{st})\mathbf{B} \mathbf{B}^H\} \right\}
    \label{eq:power}
\end{align}

The closed-form expressions of matrices $\bY_{k,k'}, \forall k,k' \in \{t, s, r\}$ are provided in \cite{williams2022electromagnetic}.
We also define $\mathbf{h}_m= \by_{RS,m}(\mathbf{Y_{s}} + \mathbf{Y_{ss}})^{-1}\mathbf{Y_{st}}$, where $\by_{RS,m}$ is the $m$-th row of  $\mathbf{Y_{RS}} = (\mathbf{Y_r} + \mathbf{Y_{rr}})^{-1} \mathbf{Y_{rs}} \in \mathbb{C}^{M\times L}$.

\textit{Notation:} Matrices/vectors are denoted by capital/small bold fonts. $(\cdot)^*$, $(\cdot)^T$, $(\cdot)^H$, and $(\cdot)^{-1}$ represent the conjugate, transpose, hermitian, and inverse operators, respectively. $\circ$ denotes the Hadamard 
product. $\bI_{M}$ is the $M \times M$ identity matrix. $\|\cdot\|$ is the spectral norm. $\text{diag}\{\bx \}$ returns a diagonal matrix in which the diagonal element are the same as $\bx$ and $\text{diag}\{\bX \}$ returns a vector consisting of the diagonal element of matrix $\bX$. $\text{tr}(\cdot)$ denotes the trace operator. $\text{Re}(\cdot)$ and $\text{Im}(\cdot)$ represent the real and imaginary part of a complex variable. $i$ is the imaginary unit and $\times$ is matrix multiplication operator.

\section{Digital beamforming}
We define a digital beamforming $\bB$ optimization problem as follows:

\begin{subequations}
\begin{align}
{\text{P1}:  \max_{\bB}}
\quad & P_{\text{tot}} \label{optaa}\\
\textrm{s.t.} \quad& \beta_d \leq P(\theta_d, \phi_d) \leq \beta_{\text{max}}, \quad d = 1,...,D \label{optbb} \\
& \gamma_m \geq \Gamma_m, \quad m=1,...,M \label{optcc} \\
& P_t \leq P_{\text{max}}, \label{optdd}
\end{align}
\end{subequations}
It should be noted that problem (P1) is nonconvex with respect to $\bB$, and our goal is to transform it into a convex problem. To achieve this, we can rewrite problem (P1) as the following convex optimization problem:
\vspace{1mm}
\framedtext{ \vspace{-1mm}
\begin{subequations}
\begin{align}
{\text{P2}:\max_{\mathcal{B}_1,...,\mathcal{B}_{M+N}}}
&\sum_{m=1}^{M+N}\text{tr}\left\{\bA_{0} \mathcal{B}_m\right\} \label{eq:P3a}  \\
\textrm{s.t.} \quad &\sum_{m=1}^{M+N}\text{tr}\{\bA_{d} \mathcal{B}_m\} \leq \beta_{\text{max}}, d = 1,...,D \label{eq:P3b}\\
&\sum_{m=1}^{M+N}\text{tr}\{\bA_{d} \mathcal{B}_m\} \geq \beta_d, d = 1,...,D \label{eq:P3c}\\
&\sum_{i=1 , i \neq m}^{M+N}\text{tr}\{\mathcal{Y}_m \mathcal{B}_i\} -\frac{1}{\Gamma_m}\text{tr}\{\mathcal{Y}_m \mathcal{B}_m\}\notag\\
&\quad\quad\quad+ \sigma^2 \leq 0, m=1,...,M  \label{eq:P3d}\\
& \sum_{m=1}^{M+N} \text{Re}\left\{\text{tr}\{\bA_P \mathcal{B}_m \}\right\} \leq P_{\text{max}}  \label{eq:P3e}\\
& \mathcal{B}_m, \succeq  0, m=1,...,M+N \label{eq:P3f}\\
& \text{rank}\{\mathcal{B}_m\} = 1, m=1,...,M+N \label{eq:P3g}
\end{align}
\end{subequations}}

\vspace{3mm}
where

\begin{align}
\mathcal{B}_m =& \mathbf{b}_m \mathbf{b}_m^H\\
\mathbf{\Psi}_m =& \by_{RSm}^H \by_{RSm}, \\
\bA_{0}=& \frac{1}{D}\mathbf{Y_{st}}^H\Tilde{\bY}^H\sum_{d=1}^D\mathbf{\Omega}_d\Tilde{\bY}\mathbf{Y_{st}} \label{eq:A0} \\
\mathbf{\Omega}_d =& \ba(\theta_d, \phi_d)^*\ba(\theta_d, \phi_d)^T, \\
\mathcal{Y}_m =& \mathbf{Y_{st}}^H\Tilde{\bY}^H \mathbf{\Psi}_m\Tilde{\bY}\mathbf{Y_{st}}\\
\bA_{d}=&\mathbf{Y_{st}}^H\Tilde{\bY}^H\mathbf{\Omega}_d\Tilde{\bY}\mathbf{Y_{st}},\\
\bA_{P}=& \frac{1}{2}(\bY_{tt} - \bY_{st}^T\Tilde{\bY}\bY_{st})
\end{align}

\subsection*{Proof:}
Problem (P1) is non-convex and challenging to solve. To make the problem convex, we need to reformulate the objective function, as well as constraints \eqref{optbb} and \eqref{optcc}. In the following subsections, we provide detailed explanations of these reformulations. The approach to rewriting \eqref{optdd} as \eqref{eq:P3e} is straightforward.

\subsection{$P_{\text{tot}}$ Reformulation}

We know \(\mathbf{B}\mathbf{B}^H = \sum_{m=1}^{M+N} \mathbf{b}_m \mathbf{b}_m^H\). By substituting this into the objective function, we have:

\begin{align}
    &\frac{1}{D}\sum_{d=1}^D \ba(\theta_d, \phi_d)^T \Tilde{\bY}\mathbf{Y_{st}} (\sum_{m=1 }^{M+N} \bb_m\bb_m^H) \mathbf{Y_{st}}^H \Tilde{\bY}^H \ba(\theta_d, \phi_d)^*\\
%    & = \frac{1}{D}\sum_{d=1}^D \text{tr}\left\{\Tilde{\bY}\mathbf{Y_{st}} (\sum_{m=1 }^{M+N} \mathcal{B}_m) \mathbf{Y_{st}}^H \Tilde{\bY}^H\ba(\theta_d, \phi_d)^*\ba(\theta_d, \phi_d)^T\right\} \label{bp1} \\
    & =\frac{1}{D}\sum_{d=1}^D \text{tr}\left\{\Tilde{\bY}\mathbf{Y_{st}} (\sum_{m=1 }^{M+N} \mathcal{B}_m) \mathbf{Y_{st}}^H \Tilde{\bY}^H\mathbf{\Omega}_d\right\} \label{bp2}\\
    & = \frac{1}{D}\text{tr}\left\{\Tilde{\bY}\mathbf{Y_{st}} (\sum_{m=1 }^{M+N} \mathcal{B}_m) \mathbf{Y_{st}}^H \Tilde{\bY}^H\sum_{d=1}^D\mathbf{\Omega}_d\right\} \label{bp3} \\
    & = \frac{1}{D}\text{tr}\left\{\mathbf{Y_{st}}^H\Tilde{\bY}^H \sum_{d=1}^D\mathbf{\Omega}_d\Tilde{\bY}\mathbf{Y_{st}} (\sum_{m=1 }^{M+N} \mathcal{B}_m)\right \} \label{bp5} \\
    & = \sum_{m=1}^{M+N}\text{tr}\left\{\bA_{0} \mathcal{B}_m\right\}
 \end{align}
where $\Tilde{\mathbf{Y}} = (\mathbf{Y}_s + \mathbf{Y}_{ss})^{-1}$.
Similarly, constraint \eqref{optbb} can be transformed to \eqref{eq:P3b} and \eqref{eq:P3c}.

\subsection{Constraint \eqref{optcc} Reformulation}
We can rewrite constraint \eqref{optcc} as follows:

\begin{align}
    & \frac{\by_{RSm}\Tilde{\bY}\mathbf{Y_{st}} \bb_m\bb_m^H \mathbf{Y_{st}}^H\Tilde{\bY}^H \by_{RSm}^H}{\by_{RSm}\Tilde{\bY}\mathbf{Y_{st}} (\sum_{i=1 , i \neq m}^{M+N} \bb_i\bb_i^H) \mathbf{Y_{st}}^H\Tilde{\bY}^H\by_{RSm}^H + \sigma^2} \geq \Gamma_m \\
    & \frac{1}{\Gamma_m}\by_{RSm}\Tilde{\bY}\mathbf{Y_{st}} \mathcal{B}_m \mathbf{Y_{st}}^H\Tilde{\bY}^H \by_{RSm}^H  \geq \notag \\
    &\quad \quad \quad \by_{RSm}\Tilde{\bY}\mathbf{Y_{st}} (\sum_{i=1 , i \neq m}^{M+N} \mathcal{B}_i) \mathbf{Y_{st}}^H\Tilde{\bY}^H\by_{RSm}^H + \sigma^2  \\
    & \frac{1}{\Gamma_m}\text{tr}\left\{\Tilde{\bY}\mathbf{Y_{st}} \mathcal{B}_m \mathbf{Y_{st}}^H\Tilde{\bY}^H \by_{RSm}^H\by_{RSm} \right\} \geq \notag \\
    & \quad\text{tr}\left\{\Tilde{\bY}\mathbf{Y_{st}} (\sum_{i=1 , i \neq m}^{M+N} \mathcal{B}_i) \mathbf{Y_{st}}^H\Tilde{\bY}^H \by_{RSm}^H\by_{RSm} \right\} + \sigma^2 \label{sinr1} \\
    & \frac{1}{\Gamma_m}\text{tr}\left\{\Tilde{\bY}\mathbf{Y_{st}} \mathcal{B}_m \mathbf{Y_{st}}^H\Tilde{\bY}^H \mathbf{\Psi}_m \right\} \geq \notag \\
    & \quad \quad \quad \quad \text{tr}\left\{\Tilde{\bY}\mathbf{Y_{st}} (\sum_{i=1 , i \neq m}^{M+N} \mathcal{B}_i) \mathbf{Y_{st}}^H\Tilde{\bY}^H \mathbf{\Psi}_m \right\} + \sigma^2 \label{sinr2}\\
    & \frac{1}{\Gamma_m}\text{tr}\left\{\mathbf{Y_{st}}^H\Tilde{\bY}^H\mathbf{\Psi}_m\Tilde{\bY}\mathbf{Y_{st}} \mathcal{B}_m  \right \} \geq \notag \\
    &\quad \quad \quad \quad \text{tr}\left\{\mathbf{Y_{st}}^H\Tilde{\bY}^H\mathbf{\Psi}_m\Tilde{\bY}\mathbf{Y_{st}} (\sum_{i=1 , i \neq m}^{M+N} \mathcal{B}_i)\right\} + \sigma^2 \label{sinr3} \\
    & \frac{1}{\Gamma_m}\text{tr}\left\{\mathcal{Y}_m \mathcal{B}_m  \right \} \geq \text{tr}\left\{\mathcal{Y}_m (\sum_{i=1 , i \neq m}^{M+N} \mathcal{B}_i)\right\} + \sigma^2 \label{sinr4}\\
    & \sum_{i=1 , i \neq m}^{M+N}\text{tr}\left\{\mathcal{Y}_m \mathcal{B}_i\right\} -\frac{1}{\Gamma_m}\text{tr}\left\{\mathcal{Y}_m \mathcal{B}_m\right\} + \sigma^2 \leq 0
\end{align}

\section{Analog Beamforming} \label{sec:Ys}
In this section, we aim to convexify problem (P3) with respect to the analog beamforming matrix $\mathbf{Y}_s$.

\begin{subequations}
\begin{align}
{\text{P3}:  \max_{\mathbf{Y}_s}}
\quad & P_{\text{tot}} \label{opta}\\
\textrm{s.t.} \quad& \beta_d \leq P(\theta_d, \phi_d) \leq \beta_{\text{max}}, \quad d = 1,...,D \label{optb} \\
& \gamma_m \geq \Gamma_m, \quad m=1,...,M \label{optc} \\
& P_t \leq P_{\text{max}}, \label{optd} \\
& \text{Re}\left\{(\bY_s)_{l,l}\right\} = R_0\geq 0, \quad l=1,...,L \label{opte} \\
& \text{Im}\{(\bY_s)_{l,l}\} \in \mathbb{R}, \quad l=1,...,L \label{optee2} 
\end{align}
\end{subequations}

We can trasnform problem (P3) to problem (P4) which is a convex problem. 
\vspace{1mm}
\framedtext{\vspace{-1mm}
\begin{subequations}
\begin{align}
{\text{P4}: \max_{\bz}} &\quad e_T +  \bw_T^H\bz^T + \bz\bw_T + \bz\bQ_T\bz^T \label{eq:P5.a}\\
\textrm{s.t.  } &\beta_d \leq e_d +  \bw_d^H\bz^T + \bz\bw_d + \bz\bQ_d\bz^T \leq \beta_{\text{max}}, \notag\\
&d = 1,...,D \label{eq:P5.b}\\
& l_m + \bs_m^H\bz^T + \bz\bs_m + \bz \bF_m \bz^T + \sigma^2  \leq 0,\notag\\
&m=1,...,M \label{eq:P5.c} \\
& \bz \bj  + \bj^H \bz^T \leq 2P_T \label{eq:P5.d}\\
&-\frac{1}{|| \widehat{\bY}_{ss}^t||} \ll \bz_l \ll \frac{1}{|| \widehat{\bY}_{ss}^t||}, l=1,...,L \label{eq:P5.e}
\end{align}
\end{subequations}}

\vspace{1mm}
where $e_T = \frac{1}{D}\sum_{d=1}^D e_d$, $\bQ_T =\frac{1}{D} \sum_{d=1}^D \bQ_d$, $\bw_T = \frac{1}{D} \sum_{d=1}^D \bw_d$, and

\begin{align}
e_d =& \text{tr}\left\{\widehat{\bY}_{ss}^t\mathbf{Y_{st}} \bB\bB^H \mathbf{Y_{st}}^H\widehat{\bY}_{ss}^{t^H}\mathbf{\Omega}_d \right\} \\
\bQ_d =& \widehat{\bY}_{ss}^t\mathbf{Y_{st}} \bB\bB^H \mathbf{Y_{st}}^H \widehat{\bY}_{ss}^{t^H} \circ \widehat{\bY}_{ss}^{t^T}\mathbf{\Omega}_d^T \widehat{\bY}_{ss}^{t^*} \\
\bW_d =& i\widehat{\bY}_{ss}^{t^H}\mathbf{\Omega}_d\widehat{\bY}_{ss}^t\mathbf{Y_{st}} \bB\bB^H \mathbf{Y_{st}}^H\widehat{\bY}_{ss}^{t^H} \\
\bw_d =& \text{diag}\left\{\bW_d \right\} \\
\bE_m =& \mathbf{Y_{st}} (\bB\bB^H -  (1 +\frac{1}{\Gamma_m})\bb_m\bb_m^H) \mathbf{Y_{st}}^H \\
l_m =& \text{tr}\left\{\widehat{\bY}_{ss}^t\bE_m \widehat{\bY}_{ss}^{t^H}\mathbf{\Psi}_m\right\} \\
\bS_m =& i\widehat{\bY}_{ss}^{t^H}\mathbf{\Psi}_m\widehat{\bY}_{ss}^t\bE_m \widehat{\bY}_{ss}^{t^H} \\
\bs_m =& \text{diag}\left\{\bS_m\right\} \\
\bF_m =& \widehat{\bY}_{ss}^t\bE_m \widehat{\bY}_{ss}^{t^H} \circ  \widehat{\bY}_{ss}^{t^T}\mathbf{\Psi}_m^T \widehat{\bY}_{ss}^{t^*} \\
\bJ =& i \widehat{\bY}_{ss}^t\bY_{st}\bB\bB^H \bY_{st}^T\widehat{\bY}_{ss}^t\\
\bj =& \text{diag}\left\{\bJ\right\} \\
P_T =& 2P_{\text{max}} -\text{Re}\left\{\text{tr}\left\{\bY_{tt}\bB\bB^H\right\} - \text{tr}\left\{\widehat{\bY}_{ss}^t \bY_{st}\bB\bB^H \bY_{st}^T\right\}\right\} 
\end{align}

\subsection*{Proof:} 
To simplify the notation and remove \eqref{opte}, we define $\Tilde{\bY}_{ss} = \bY_{ss} + \text{Re}\{\bY_s\}$ and $\text{Im}\{\bY_s\}  = \bY$.
To solve (P3), we use the Neumann series approximation for $(\Tilde{\bY}_{ss} + i\bY)^{-1}$.
Consequently, we update $\bY$ iteratively as $\bY^{t+1} = \bY^{t} + \Delta$, where $\Delta$ is a $L\times L$ diagonal matrix used as small increments, and $t$ shows the iteration step.
Under the considered case of interest, in which $(\Tilde{\bY}_{ss} + i\bY^t)$ is invertible, we approximate $(\Tilde{\bY}_{ss} + i\bY^{t+1})^{-1}$ using Neuman series as:

\begin{align}
    (\Tilde{\bY}_{ss} + i\bY^{t+1})^{-1} \approx \widehat{\bY}_{ss}^t - i\widehat{\bY}_{ss}^t \Delta \widehat{\bY}_{ss}^t
    \label{eq:Neuman}
\end{align}
where $\widehat{\bY}_{ss}^t = (\Tilde{\bY}_{ss} + i\bY^t)^{-1}$.
To ensure the approximation is accurate enough, the condition $||\Delta \widehat{\bY}_{ss}^t|| \ll 1$ needs to hold \cite[Eq. (4.17)]{stewart1998matrix}. Using the fact that $||\Delta \widehat{\bY}_{ss}^t|| \leq ||\Delta|| ||\widehat{\bY}_{ss}^t||$, the inequality $||\Delta \widehat{\bY}_{ss}^t|| \ll 1$ is equivalent to $||\Delta|| \ll \frac{1}{|| \widehat{\bY}_{ss}^t||}$. 
We then substitute the approximation of $(\bY_s + \widehat{\bY}_{ss})^{-1}$ into (P3), where the new optimization variable is $\Delta$. Letting $\bz = \text{diag}\{\Delta\} \in \mathbb{R}^{L \times 1}$, we explain in the following subsections how each function in (P4) is transformed.

\subsection{$P_{\text{tot}}$ Reformulation}
For the summation of beampattern gains in the objective function, we replace \eqref{eq:Neuman} in \eqref{eq:BPtot} as follows:
\vspace{-2mm}
\begin{align}
     \frac{1}{D}& \sum_{d=1}^D \ba(\theta_d, \phi_d)^T (\widehat{\bY}_{ss}^t - i\widehat{\bY}_{ss}^t \Delta \widehat{\bY}_{ss}^t)\mathbf{Y_{st}} \bB\bB^H \mathbf{Y_{st}}^H (\widehat{\bY}_{ss}^{t^H} \notag\\
    &\quad \quad \quad \quad \quad \quad \quad \quad \quad \quad + i\widehat{\bY}_{ss}^{t^H} \Delta \widehat{\bY}_{ss}^{t^H}) \ba(\theta_d, \phi_d)^* \\
    = & \frac{1}{D}\sum_{d=1}^D \text{tr} \biggl\{(\widehat{\bY}_{ss}^t - i\widehat{\bY}_{ss}^t \Delta \widehat{\bY}_{ss}^t)\mathbf{Y_{st}} \bB\bB^H \mathbf{Y_{st}}^H (\widehat{\bY}_{ss}^{t^H} \notag\\
    &\quad \quad \quad \quad \quad + i\widehat{\bY}_{ss}^{t^H} \Delta \widehat{\bY}_{ss}^{t^H})\ba(\theta_d, \phi_d)^*\ba(\theta_d, \phi_d)^T\biggr \}\label{bpp1}\\
    = & \frac{1}{D}\sum_{d=1}^D \text{tr}\biggl\{(\widehat{\bY}_{ss}^t - i\widehat{\bY}_{ss}^t \Delta \widehat{\bY}_{ss}^t)\mathbf{Y_{st}} \bB\bB^H \mathbf{Y_{st}}^H (\widehat{\bY}_{ss}^{t^H} \notag\\
    &\quad \quad \quad \quad \quad \quad \quad \quad \quad \quad \quad \quad + i\widehat{\bY}_{ss}^{t^H} \Delta \widehat{\bY}_{ss}^{t^H})\mathbf{\Omega}_d\biggr\} \label{bpp2}\\
    = & \frac{1}{D}\sum_{d=1}^D \text{tr}\biggl\{(\widehat{\bY}_{ss}^t\mathbf{Y_{st}} \bB\bB^H \mathbf{Y_{st}}^H - i\widehat{\bY}_{ss}^t \Delta \widehat{\bY}_{ss}^t\mathbf{Y_{st}} \bB\bB^H\notag\\
    &\quad \quad \quad \quad\times \mathbf{Y_{st}}^H) (\widehat{\bY}_{ss}^{t^H}\mathbf{\Omega}_d + i\widehat{\bY}_{ss}^{t^H} \Delta \widehat{\bY}_{ss}^{t^H}\mathbf{\Omega}_d)\biggr\}\label{bpp3} \\
    = & \frac{1}{D}\sum_{d=1}^D \biggl[\text{tr}\left\{\widehat{\bY}_{ss}^t\mathbf{Y_{st}} \bB\bB^H \mathbf{Y_{st}}^H\widehat{\bY}_{ss}^{t^H}\mathbf{\Omega}_d \right\} \notag\\
    &\quad +  i\text{tr}\left\{\widehat{\bY}_{ss}^t\mathbf{Y_{st}} \bB\bB^H \mathbf{Y_{st}}^H\widehat{\bY}_{ss}^{t^H} \Delta \widehat{\bY}_{ss}^{t^H}\mathbf{\Omega}_d \right\} \notag \\
    &\quad   -  i\text{tr}\left\{ \widehat{\bY}_{ss}^t \Delta \widehat{\bY}_{ss}^t\mathbf{Y_{st}} \bB\bB^H \mathbf{Y_{st}}^H \widehat{\bY}_{ss}^{t^H}\mathbf{\Omega}_d\right\} \notag \\
    &\quad  + \text{tr}\left\{\widehat{\bY}_{ss}^t \Delta \widehat{\bY}_{ss}^t\mathbf{Y_{st}} \bB\bB^H \mathbf{Y_{st}}^H \widehat{\bY}_{ss}^{t^H} \Delta \widehat{\bY}_{ss}^{t^H}\mathbf{\Omega}_d\right\}\biggr] \label{bpp4}\\
    = & \frac{1}{D}\sum_{d=1}^D \biggl[e_d +  i\text{tr}\left\{\Delta\widehat{\bY}_{ss}^{t^H}\mathbf{\Omega}_d\widehat{\bY}_{ss}^t\mathbf{Y_{st}} \bB\bB^H \mathbf{Y_{st}}^H\widehat{\bY}_{ss}^{t^H} \right\} \notag\\
    &\quad -  i\text{tr}\left\{ \widehat{\bY}_{ss}^t\mathbf{Y_{st}} \bB\bB^H \mathbf{Y_{st}}^H \widehat{\bY}_{ss}^{t^H}\mathbf{\Omega}_d \widehat{\bY}_{ss}^t \Delta\right\}\notag \\
    &\quad +  \text{tr}\left\{ \Delta \widehat{\bY}_{ss}^t\mathbf{Y_{st}} \bB\bB^H \mathbf{Y_{st}}^H \widehat{\bY}_{ss}^{t^H} \Delta \widehat{\bY}_{ss}^{t^H}\mathbf{\Omega}_d \widehat{\bY}_{ss}^t\right\}\biggr] \label{bpp5}\\
    = & \frac{1}{D}\sum_{d=1}^D \biggl[e_d +  \text{tr}\left\{ \Delta \bW_d\right\} +  \text{tr}\left\{\bW_d^H \Delta \right\} \notag\\
    &\quad + \text{tr}\left\{ \Delta \widehat{\bY}_{ss}^t\mathbf{Y_{st}} \bB\bB^H \mathbf{Y_{st}}^H \widehat{\bY}_{ss}^{t^H} \Delta \widehat{\bY}_{ss}^{t^H}\mathbf{\Omega}_d \widehat{\bY}_{ss}^t\right\}\biggr] \label{bpp6}
    \end{align}
    \begin{align}
    = & \frac{1}{D}\sum_{d=1}^D \biggl[e_d + \bz\bw_d  +  \bw_d^H\bz^T \notag\\
    &\quad  + \bz(\widehat{\bY}_{ss}^t\mathbf{Y_{st}} \bB\bB^H \mathbf{Y_{st}}^H \widehat{\bY}_{ss}^{t^H} \circ \widehat{\bY}_{ss}^{t^T}\mathbf{\Omega}_d^T \widehat{\bY}_{ss}^{t^*})\bz^T\biggr] \label{bpp7} \\
    = & \frac{1}{D}\sum_{d=1}^D \biggl[e_d + \bz\bw_d +  \bw_d^H\bz^T + \bz\bQ_d\bz^T\biggr] \label{bpp8}\\
    = & e_T + \bz\bw_T +  \bw_T^H\bz^T + \bz\bQ_T\bz^T \label{bpp9}
\end{align}

The beampattern gains in constraint \eqref{optb} can be reformulated as \eqref{eq:P5.b} in a similar way.

\subsection{Constraint \eqref{optc} Reformulation}
By replacing \eqref{eq:Neuman} in \eqref{optb}, we can rewrite it as follows:

\begin{align}
& \frac{1}{\Gamma_m} \by_{RSm}(\widehat{\bY}_{ss}^t - i\widehat{\bY}_{ss}^t \Delta \widehat{\bY}_{ss}^t)\mathbf{Y_{st}} \bb_m\bb_m^H \mathbf{Y_{st}}^H(\widehat{\bY}_{ss}^{t^H} \notag \\
&+ i\widehat{\bY}_{ss}^{t^H} \Delta \widehat{\bY}_{ss}^{t^H})\by_{RSm}^H \geq \by_{RSm}(\widehat{\bY}_{ss}^t - i\widehat{\bY}_{ss}^t \Delta \widehat{\bY}_{ss}^t)\mathbf{Y_{st}} \bR_m \notag\\
&\quad \quad \quad \quad \quad \times \mathbf{Y_{st}}^H(\widehat{\bY}_{ss}^{t^H} + i\widehat{\bY}_{ss}^{t^H} \Delta \widehat{\bY}_{ss}^{t^H})\by_{RSm}^H + \sigma^2\label{ss1} \\
& \by_{RSm}(\widehat{\bY}_{ss}^t - i\widehat{\bY}_{ss}^t \Delta \widehat{\bY}_{ss}^t)\mathbf{Y_{st}} (\bR_m -  \frac{1}{\Gamma_m}\bb_m\bb_m^H) \mathbf{Y_{st}}^H(\widehat{\bY}_{ss}^{t^H} \notag\\
&\quad \quad \quad \quad \quad \quad \quad \quad \quad + i\widehat{\bY}_{ss}^{t^H} \Delta \widehat{\bY}_{ss}^{t^H})\by_{RSm}^H + \sigma^2 \leq 0 \label{ss2}\\
& \text{tr}\biggl\{(\widehat{\bY}_{ss}^t - i\widehat{\bY}_{ss}^t \Delta \widehat{\bY}_{ss}^t)\mathbf{Y_{st}} (\bR_m -  \frac{1}{\Gamma_m}\bb_m\bb_m^H) \mathbf{Y_{st}}^H(\widehat{\bY}_{ss}^{t^H} \notag \\
&\quad \quad \quad \quad \quad \quad + i\widehat{\bY}_{ss}^{t^H} \Delta \widehat{\bY}_{ss}^{t^H}) \by_{RSm}^H \by_{RSm}\biggr\} + \sigma^2 \leq 0  \label{ss3}\\
& \text{tr}\biggl\{(\widehat{\bY}_{ss}^t - i\widehat{\bY}_{ss}^t \Delta \widehat{\bY}_{ss}^t)\mathbf{Y_{st}} (\bR_m -  \frac{1}{\Gamma_m}\bb_m\bb_m^H) \mathbf{Y_{st}}^H (\widehat{\bY}_{ss}^{t^H} \notag\\
&\quad \quad \quad \quad \quad \quad \quad \quad \quad+ i\widehat{\bY}_{ss}^{t^H} \Delta \widehat{\bY}_{ss}^{t^H}) \mathbf{\Psi}_m\biggr\} + \sigma^2 \leq 0 \label{ss4}\\
& \text{tr}\biggl\{(\widehat{\bY}_{ss}^t - i\widehat{\bY}_{ss}^t \Delta \widehat{\bY}_{ss}^t)\bE_m (\widehat{\bY}_{ss}^{t^H} + i\widehat{\bY}_{ss}^{t^H} \Delta \widehat{\bY}_{ss}^{t^H}) \mathbf{\Psi}_m\biggr\} \notag \\
&\quad \quad \quad \quad \quad \quad \quad \quad \quad \quad \quad \quad \quad \quad \quad \quad \quad \quad+ \sigma^2 \leq 0 \label{ss5}\\
& \text{tr}\biggl\{(\widehat{\bY}_{ss}^t\bE_m - i\widehat{\bY}_{ss}^t \Delta \widehat{\bY}_{ss}^t\bE_m) (\widehat{\bY}_{ss}^{t^H}\mathbf{\Psi}_m \notag \\
& \quad \quad \quad \quad \quad \quad \quad \quad \quad+ i\widehat{\bY}_{ss}^{t^H} \Delta \widehat{\bY}_{ss}^{t^H}\mathbf{\Psi}_m) \biggr\} + \sigma^2 \leq 0 \label{ss6}\\
& \text{tr}\left\{\widehat{\bY}_{ss}^t\bE_m \widehat{\bY}_{ss}^{t^H}\mathbf{\Psi}_m\right\} + i\text{tr}\left\{\widehat{\bY}_{ss}^t\bE_m \widehat{\bY}_{ss}^{t^H} \Delta \widehat{\bY}_{ss}^{t^H}\mathbf{\Psi}_m\right\} \notag \\
&\quad \quad \quad - i\text{tr}\left\{\widehat{\bY}_{ss}^t \Delta \widehat{\bY}_{ss}^t\bE_m  \widehat{\bY}_{ss}^{t^H}\mathbf{\Psi}_m\right\} \notag\\
&\quad \quad \quad + \text{tr}\left\{\widehat{\bY}_{ss}^t \Delta \widehat{\bY}_{ss}^t\bE_m \widehat{\bY}_{ss}^{t^H} \Delta \widehat{\bY}_{ss}^{t^H}\mathbf{\Psi}_m\right\} + \sigma^2  \leq 0 \label{ss7} \\
 & l_m  + i \text{tr}\{ \Delta\widehat{\bY}_{ss}^{t^H}\mathbf{\Psi}_m\widehat{\bY}_{ss}^t\bE_m \widehat{\bY}_{ss}^{t^H} \} \notag \\
 &\quad \quad \quad - i\text{tr}\{ \widehat{\bY}_{ss}^t\bE_m  \widehat{\bY}_{ss}^{t^H}\mathbf{\Psi}_m\widehat{\bY}_{ss}^t \Delta \} \notag\\
 &\quad \quad \quad + \text{tr}\{ \Delta \widehat{\bY}_{ss}^t\bE_m \widehat{\bY}_{ss}^{t^H} \Delta \widehat{\bY}_{ss}^{t^H}\mathbf{\Psi}_m \widehat{\bY}_{ss}^t\} + \sigma^2  \leq 0 \label{ss8}\\
 & l_m + \text{tr}\left\{ \Delta \bS_m\right\}  + \text{tr}\left\{\bS_m^H \Delta \right\} \notag \\
 &\quad \quad \quad + \bz (\widehat{\bY}_{ss}^t\bE_m \widehat{\bY}_{ss}^{t^H} \circ  \widehat{\bY}_{ss}^{t^T}\mathbf{\Psi}_m^T \widehat{\bY}_{ss}^{t^*}) \bz^T + \sigma^2  \leq 0 \label{ss9} \\
 & l_m + \bz\bs_m + \bs_m^H\bz^T + \bz \bF_m \bz^T + \sigma^2  \leq 0 \label{ss10}
\end{align}
where $\bR_m = \bB\bB^H - \bb_m\bb_m^H$.

\subsection{Constraint \eqref{optd} Reformulation}
The power constraint, after replacing \eqref{eq:Neuman} in \eqref{eq:power}, can be rewritten as:

\begin{align}
    &\frac{1}{2} \text{Re}\biggl\{\text{tr}\{(\bY_{tt} - \bY_{st}^T(\widehat{\bY}_{ss}^t \notag \\
    &\quad \quad \quad \quad \quad \quad \quad  -i\widehat{\bY}_{ss}^t \Delta \widehat{\bY}_{ss}^t)\bY_{st})\bB\bB^H \}\biggr\} \leq P_{\text{max}} \\
    &- \text{Re}\left\{\text{tr}\{\bY_{st}^T(\widehat{\bY}_{ss}^t - i\widehat{\bY}_{ss}^t \Delta \widehat{\bY}_{ss}^t)\bY_{st}\bB\bB^H \}\right\} \leq  \notag\\
    &\quad \quad \quad \quad \quad \quad \quad \quad \quad \quad \;\; 2P_{\text{max}} - \text{Re}\left\{\text{tr}\{\bY_{tt}\bB\bB^H\}\right\} \label{pw1} \\
    &- \text{Re}\left\{\text{tr}\{(\widehat{\bY}_{ss}^t - i\widehat{\bY}_{ss}^t \Delta \widehat{\bY}_{ss}^t)\bY_{st}\bB\bB^H \bY_{st}^T\}\right\} \leq 2P_{\text{max}}\notag \\
    &\quad \quad \quad \quad \quad \quad \quad \quad \quad \quad \quad \quad \quad - \text{Re}\left\{\text{tr}\{\bY_{tt}\bB\bB^H\}\right\} \label{pw2} \\
    &\text{Re}\left\{\text{tr}\{ i\widehat{\bY}_{ss}^t \Delta \widehat{\bY}_{ss}^t\bY_{st}\bB\bB^H \bY_{st}^T\}\right\} \leq 2P_{\text{max}} 
    \notag \\
    &\quad - \text{Re}\left\{\text{tr}\{\bY_{tt}\bB\bB^H\}\right\} -\text{Re}\left\{\text{tr}\{\widehat{\bY}_{ss}^t \bY_{st}\bB\bB^H \bY_{st}^T\}\right\} \label{pw3}\\
    &  \text{Re}\left\{\text{tr}\{ i \Delta \widehat{\bY}_{ss}^t\bY_{st}\bB\bB^H \bY_{st}^T\widehat{\bY}_{ss}^t\}\right\} \leq P_T \label{pw4}\\
    &  \text{Re}\left\{\text{tr}\{ \Delta \bJ \}\right\} \leq P_T \label{pw5}\\
    & \text{tr}\{ \Delta \bJ \} + \text{tr}\{  \bJ^H \Delta\}\leq 2P_T  \label{pw6}\\
    & \bz \bj  + \bj^H \bz^T \leq 2P_T \label{pw7}
\end{align}

\section{Proof of objective function being upper bounded}
We can rewrite $P_{\text{tot}}$ as:

\begin{align}
  P_{\text{tot}}  &= \text{tr}\left\{\frac{1}{D}\mathbf{Y_{st}}^H\Tilde{\bY}^H\sum_{d=1}^D\mathbf{\Omega}_d\Tilde{\bY}\mathbf{Y_{st}} \bB\bB^H\right\} \\
  &= \text{tr}\left\{\bB^H \mathbf{Y_{st}}^H \Tilde{\bY}^H\mathbf{\Omega}_T\Tilde{\bY}\mathbf{Y_{st}} \bB\right\} \label{eq:pd}
\end{align}
where $\mathbf{\Omega}_T = \frac{1}{D}\sum_{d=1}^D\mathbf{\Omega}_d$. Since $\bB^H \mathbf{Y_{st}}^H \Tilde{\bY}^H\mathbf{\Omega}_T\Tilde{\bY}\mathbf{Y_{st}} \bB$ in \eqref{eq:pd} is \gls{PSD} matrix, we have:

\begin{align}
   &\text{tr}\{\bB^H \mathbf{Y_{st}}^H \Tilde{\bY}^H\mathbf{\Omega}_T\Tilde{\bY}\mathbf{Y_{st}} \bB\} \\
   &= \text{tr}\{\Tilde{\bY}^H\mathbf{\Omega}_T\Tilde{\bY}\mathbf{Y_{st}} \bB\bB^H\mathbf{Y_{st}}^H\}  \\
    & \leq || \Tilde{\bY}^H\mathbf{\Omega}_T\Tilde{\bY}\mathbf{Y_{st}} \bB\bB^H\mathbf{Y_{st}}^H || \times \text{min}\{ L,M,N \} \\
    & \leq || \Tilde{\bY}^H ||. ||\mathbf{\Omega}_T || . || \Tilde{\bY} || . || \mathbf{Y_{st}} \bB\bB^H\mathbf{Y_{st}}^H || \times \text{min}\{ L,M,N \} \label{eq:p4}\\
    & \leq || \Tilde{\bY}^H || . || \Tilde{\bY} || . || \mathbf{Y_{st}} \bB\bB^H\mathbf{Y_{st}}^H || \times L\times \text{min}\{ L,M,N \} \label{eq:p5}
\end{align}

In \eqref{eq:p4}, we have $||\mathbf{\Omega}_T || \leq \frac{1}{D}\sum_{d=1}^D ||\mathbf{\Omega}_d || \leq L$ because $|\mathbf{\Omega}_d|_{l,l'} \leq 1, \forall l,l'= 1,...,L$. Since $\bY_s + \mathbf{Y_{ss}}$ is 
invertible in the case study of interest, then $|| \Tilde{\bY} ||_2$ in \eqref{eq:p5} is upper bounded by the smallest singular value of $\bY_s + \mathbf{Y_{ss}}$. A similar justification applies for $\Tilde{\bY}^H$. Eventually, $|| \mathbf{Y_{st}} \bB\bB^H\mathbf{Y_{st}}^H ||$ in \eqref{eq:p5} can be concluded that is upper bounded using the power constraint. To prove it, we rewrite \eqref{optd} as:

\begin{align}
    &\text{Re}\left\{\text{tr}\{\bY_{tt}\mathbf{B} \mathbf{B}^H\}\right\} \notag \\
    &- \text{Re}\left\{\text{tr}\{\bY_{st}^T(\bY_s + \bY_{ss})^{ - 1}\bY_{st})\mathbf{B} \mathbf{B}^H \}\right\} \leq 2P_{\text{max}} 
    \label{eq:power2}
\end{align}
In \eqref{eq:power2}, $\bY_{tt}$ is a diagonal matrix, assuming that the \gls{RF} chains are isolated from each other, and its diagonal elements are all imaginary \cite[Eq. 36]{williams2022electromagnetic}. This implies that, $\bY_{tt} = \bY_{tt}^T$ and $\bY_{tt}^H = -\bY_{tt}$. Therefore, we can rewrite the first term as follows:

\begin{align}
    &\text{Re}\left\{\text{tr}\{\bY_{tt}\mathbf{B} \mathbf{B}^H\}\right\} = \frac{1}{2}(\text{tr}\{\bY_{tt}\mathbf{B} \mathbf{B}^H\} + \text{tr}\{\mathbf{B} \mathbf{B}^H \bY_{tt}^H \}) \\
    & = \frac{1}{2}(\text{tr}\{\bY_{tt}\mathbf{B} \mathbf{B}^H\} + \text{tr}\{\bY_{tt}^H \mathbf{B} \mathbf{B}^H \}) \\
    & = \frac{1}{2}(\text{tr}\{\bY_{tt}\mathbf{B} \mathbf{B}^H\} - \text{tr}\{\bY_{tt} \mathbf{B} \mathbf{B}^H \}) = 0
\end{align}

Thus, \eqref{eq:power2} can be written as:

\begin{align}
    &- \text{Re}\left\{\text{tr}\{\bY_{st}^T(\bY_s + \bY_{ss})^{ - 1}\bY_{st})\mathbf{B} \mathbf{B}^H \}\right\} \leq 2P_{\text{max}}\\
    & - \text{Re}\left\{\text{tr}\{\Tilde{\bY} \bY_{st}\mathbf{B} \mathbf{B}^H \bY_{st}^T\}\right\} \leq 2P_{\text{max}} \label{eq:power3}
\end{align}
where we replaced $(\bY_s + \bY_{ss})^{ - 1}$ by $\Tilde{\bY}$. Since $\bY_{ss}$ is symmetric and $\bY_s$ is a diagonal matrix, $\bY_s + \bY_{ss}$ is also symmetric, and its inverse, i.e., $\Tilde{\bY}$, is symmetric as well.
We denote $\Tilde{\bY}_{T} = -\bY_{st}\mathbf{B} \mathbf{B}^H \bY_{st}^T$, which is a Hermitian matrix, and replace it in \eqref{eq:power3}.

\begin{align}
    &\text{Re}\left\{\text{tr}\{\Tilde{\bY} \Tilde{\bY}_{T}\}\right\} \leq 2P_{\text{max}}\\
    &\text{tr}\{\Tilde{\bY} \Tilde{\bY}_{T}\} + \text{tr}\{\Tilde{\bY}_{T}^H  \Tilde{\bY}^H \} \leq 4P_{\text{max}}\\
    &\text{tr}\{(\Tilde{\bY} + \Tilde{\bY}^H) \Tilde{\bY}_{T}\} \leq 4P_{\text{max}}\\
    & || (\Tilde{\bY} + \Tilde{\bY}^H) \Tilde{\bY}_{T} ||_2 \leq 4 P_{\text{max}} \label{eq:power4}
\end{align}

In \eqref{eq:power4}, we use $ || (\Tilde{\bY} + \Tilde{\bY}^H) \Tilde{\bY}_{T} || \leq \text{tr}\{(\Tilde{\bY} + \Tilde{\bY}^H) \Tilde{\bY}_{T}\} $.
According to \eqref{eq:Neuman}, we approximate $\Tilde{\bY}$ as $\Tilde{\bY} \approx \widehat{\bY}_{ss} - i\widehat{\bY}_{ss} \Delta \widehat{\bY}_{ss}$ in each iteration. Based on the condition for an accurate approximation, we have:

\begin{align}
    &||i\Delta \widehat{\bY}_{ss}|| \ll 1 \\
    &||\widehat{\bY}_{ss}|| . ||i\Delta \widehat{\bY}_{ss}|| \ll ||\widehat{\bY}_{ss}||
\end{align}
Since $||i\widehat{\bY}_{ss} \Delta \widehat{\bY}_{ss}|| \leq ||\widehat{\bY}_{ss}|| . ||i\Delta \widehat{\bY}_{ss}||$, we have:

\begin{align}
    ||i\widehat{\bY}_{ss} \Delta \widehat{\bY}_{ss}|| \ll ||\widehat{\bY}_{ss}|| 
    \label{eq:inequality1}
\end{align}

Using the triangle inequality, we have:
\begin{equation}
    || \widehat{\bY}_{ss} - i\widehat{\bY}_{ss} \Delta \widehat{\bY}_{ss} || \leq || \widehat{\bY}_{ss} || + || i\widehat{\bY}_{ss} \Delta \widehat{\bY}_{ss} ||
\end{equation}

Based on \eqref{eq:inequality1}, we know that $||i\widehat{\bY}_{ss} \Delta \widehat{\bY}_{ss}|| \ll ||\widehat{\bY}_{ss}|| + || i\widehat{\bY}_{ss} \Delta \widehat{\bY}_{ss} ||$. Therefore, we can conclude the following:

\begin{align}
   &|| i\widehat{\bY}_{ss} \Delta \widehat{\bY}_{ss} || \ll || \widehat{\bY}_{ss} - i\widehat{\bY}_{ss} \Delta \widehat{\bY}_{ss} || \approx ||\Tilde{\bY}||
   \label{eq:inequality2}\\
   &(|| i\widehat{\bY}_{ss} \Delta \widehat{\bY}_{ss} || + || -i\widehat{\bY}_{ss}^H \Delta \widehat{\bY}_{ss}^H ||) . || \Tilde{\bY}_{T} || \ll (||\Tilde{\bY}|| \notag \\
   &\quad \quad \quad \quad \quad \quad \quad \quad \quad \quad \quad \quad \quad \quad \quad+ ||\Tilde{\bY}^H || ) . || \Tilde{\bY}_{T} ||
\end{align}

Using $|| (\Tilde{\bY} + \Tilde{\bY}^H) \Tilde{\bY}_{T} || \leq (|| \Tilde{\bY} || +||\Tilde{\bY}^H ||). ||\Tilde{\bY}_{T} || $, we derive the following:

\begin{align}
    &(|| i\widehat{\bY}_{ss} \Delta \widehat{\bY}_{ss} || + || -i\widehat{\bY}_{ss}^H \Delta \widehat{\bY}_{ss}^H ||) . || \Tilde{\bY}_{T} || \ll  \notag \\
    &\quad \quad \quad \quad \quad \quad \quad \quad \quad \quad || (\Tilde{\bY} + \Tilde{\bY}^H) \Tilde{\bY}_{T} || \leq 4 P_{\text{max}}
    \label{eq:inequality4}
\end{align}

Next, we need to prove that $|| i\widehat{\bY}_{ss} \Delta \widehat{\bY}_{ss} ||$  is upper bounded. To do so, we have the following:

\begin{align}
    || i\widehat{\bY}_{ss} \Delta \widehat{\bY}_{ss} || \leq || \widehat{\bY}_{ss} || . || i\Delta || . || \widehat{\bY}_{ss} || = || \widehat{\bY}_{ss} ||^2 . || \Delta || 
    \label{eq:inequality3}
\end{align}

In \eqref{eq:inequality3}, $\widehat{\bY}_{ss}$, as defined in Section \ref{sec:Ys}, is upper bounded by the smallest singular value of $\Tilde{\bY}_{ss} + i\bY$. Additionally, $|| \Delta ||$ is upper bounded as explained in \ref{sec:Ys}. Therefore, $|| \widehat{\bY}_{ss} ||^2 . || \Delta || $ and consequently $|| i\widehat{\bY}_{ss} \Delta \widehat{\bY}_{ss} || $ are upper bounded. Similarly, we can prove $|| -i\widehat{\bY}_{ss}^H \Delta \widehat{\bY}_{ss}^H ||$ is also upper bounded. Therefore, $(|| i\widehat{\bY}_{ss} \Delta \widehat{\bY}_{ss} || + || -i\widehat{\bY}_{ss}^H \Delta \widehat{\bY}_{ss}^H ||)$ in \eqref{eq:inequality4} is upper bounded. Since $(|| i\widehat{\bY}_{ss} \Delta \widehat{\bY}_{ss} || + || -i\widehat{\bY}_{ss}^H \Delta \widehat{\bY}_{ss}^H ||) . || \Tilde{\bY}_{T} ||$ is less than $4 P_{\text{max}}$, and $(|| i\widehat{\bY}_{ss} \Delta \widehat{\bY}_{ss} || + || -i\widehat{\bY}_{ss}^H \Delta \widehat{\bY}_{ss}^H ||)$ is upper bounded, we can conclude $|| \Tilde{\bY}_{T} ||$ is also upper bounded. All elements of $\bY_{st}$ in $\Tilde{\bY}_{T}$ are imaginary \cite[Eq. 32]{williams2022electromagnetic}. Thus, $-\bY_{st}^T = \bY_{st}^H$, and consequently, $|| \Tilde{\bY}_{T} || = || \mathbf{Y_{st}} \bB\bB^H\mathbf{Y_{st}}^H ||$, which has been proven to be upper bounded. This concludes the proof.

\section{Conclusion}
This document serves as supplementary material for a journal paper submission, providing comprehensive mathematical proofs and derivations that underpin the results presented in the main manuscript. In this study, we formulate an \gls{ISAC} optimization problem that seeks to equitably maximize beampattern gains across multiple directions of interest, while simultaneously satisfying the \gls{SINR} requirements for communication users and adhering to a total power constraint. To address this problem, we introduce a dedicated optimization algorithm and rigorously prove its convergence.

\bibliographystyle{ieeetr}
\bibliography{ref}

\begin{thebibliography}{1}

\bibitem{smith2017analysis}
D.~R. Smith, O.~Yurduseven, L.~P. Mancera, P.~Bowen, and N.~B. Kundtz, ``Analysis of a waveguide-fed metasurface antenna,'' {\em Physical Review Applied}, vol.~8, no.~5, p.~054048, 2017.

\bibitem{shlezinger2021dynamic}
N.~Shlezinger {\em et~al.}, ``Dynamic metasurface antennas for {6G} extreme massive mimo communications,'' {\em IEEE Wireless Communications}, vol.~28, no.~2, pp.~106--113, 2021.

\bibitem{kimaryo2022downlink}
S.~F. Kimaryo and K.~Lee, ``Downlink beamforming for dynamic metasurface antennas,'' {\em IEEE Transactions on Wireless Communications}, vol.~22, no.~7, pp.~4745--4755, 2022.

\bibitem{shlezinger2019dynamic}
N.~Shlezinger {\em et~al.}, ``Dynamic metasurface antennas for uplink massive mimo systems,'' {\em IEEE transactions on communications}, vol.~67, no.~10, pp.~6829--6843, 2019.

\bibitem{williams2022electromagnetic}
R.~J. Williams, P.~Ram{\'\i}rez-Espinosa, J.~Yuan, and E.~De~Carvalho, ``Electromagnetic based communication model for dynamic metasurface antennas,'' {\em IEEE Transactions on Wireless Communications}, 2022.

\bibitem{stewart1998matrix}
G.~W. Stewart, {\em Matrix algorithms: volume 1: basic decompositions}.
\newblock SIAM, 1998.

\end{thebibliography}

\end{document}